# Fine Grid Numerical Solutions of Triangular Cavity Flow

Ercan Erturk[1] and Orhan Gokcol[2]

[1] Gebze Institute of Technology, Energy Systems Engineering Department, Gebze, Kocaeli 41400, Turkey
[2] Bahcesehir University, Computer Engineering Department, Besiktas, Istanbul 34349, Turkey

**Abstract.** Numerical solutions of 2-D steady incompressible flow inside a triangular cavity are presented. For the purpose of comparing our results with several different triangular cavity studies with different triangle geometries, a general triangle mapped onto a computational domain is considered. The Navier-Stokes equations in general curvilinear coordinates in streamfunction and vorticity formulation are numerically solved. Using a very fine grid mesh, the triangular cavity flow is solved for high Reynolds numbers. The results are compared with the numerical solutions found in the literature and also with analytical solutions as well. Detailed results are presented.

## 1 INTRODUCTION

Flows inside closed geometries have always been the focus of attention of Computational Fluid Dynamics (CFD) studies. As an example, the square driven cavity flow can be counted. Most probably the interest to the driven cavity flow is due to the simplicity of the geometry and also the simplicity of the application of the boundary conditions. Despite its simplicity, the driven cavity flow shows interesting flow features as the Reynolds number increases.

In the literature it is possible to find many numerical studies on the driven cavity flow. Among numerous papers found in the literature, Erturk et al. [6], Erturk and Gokcol [7], Barragy and Carey [1], Botella and Peyret [4], Rubin and Khosla [18], Benjamin and Denny [3], Ghia et al. [9] and Li et al. [12] are examples of numerical studies on the driven cavity flow. Burggraf [5] applied the Batchelor's [2] model to driven cavity flow and obtained the theoretical core vorticity value at infinite Reynolds number. Erturk et al. [6] introduced a new efficient numerical method and with this method, using a fine grid mesh, they solved the driven cavity flow at very high Reynolds numbers up to $Re$=21000. They showed that at high Reynolds number the vorticity at the center of the primary vortex asymptotically approaches to a value close to the theoretical value obtained by Burggraf [5]. The difference between the numerical values of Erturk et al. [6] and the theoretical value can be attributed to the thin boundary layer developed along the solid walls and the counter rotating vortices appear at the corners as the Reynolds number increases.

The 2-D steady incompressible flow inside a triangular driven cavity is also an interesting subject like the square driven cavity flow. This flow was studied numerically by McQuain et al. [14], Ribbens et al. [17], Jyotsna and Vanka [11], Li and Tang [13] and Gaskell et al. [8]. After a brief literature survey, to the best of our knowledge, these are the only numerical studies found in the literature on the steady incompressible flow inside a triangular driven cavity. The results of these studies, however, show some discrepancies. This constitutes the main motivation of this study.

Apart from numerical studies, Moffatt [15] studied the triangular cavity flow analytically in the Stokes regime. Moffatt [15] showed that the intensities of eddies and the distance of eddy centers from the corner, follow a geometric sequence. Moffatt's [15] results will provide a mathematical check on our numerical results, as it was also used in Jyotsna and Vanka's [11] and Gaskell et al.'s [8] study.

McQuain et al. [14] applied the Batchelor's [2] mean square law to triangular cavity flow and analytically obtained the inviscid core vorticity for infinite Reynolds number. In this study we will discuss the predictions of the mean square law for the triangular cavity flow and its applicability for the triangular cavity flow.

Erturk et al. [6] stated that for square driven cavity flow fine grid mesh is necessary for the reliability of the solution when high Reynolds numbers are considered. In order to obtain high accurate numerical solutions, follow-

*Correspondence to*: ercanerturk@gyte.edu.tr
*URL*: http://www.cavityflow.com

ing Erturk *et al.* [6], we will use a fine grid mesh with $(512 \times 512)/2$ grid points.

The aim of this study is then, to solve the 2-D steady incompressible triangular cavity flow using a fine grid mesh and present accurate solutions. We will compare our results in detail with the numerical studies found in the literature. We will also compare our Stokes regime solutions with analytical results of Moffatt [15] and our high Reynolds number solutions with Batchelor's [2] model which consists of an inviscid core with uniform vorticity coupled to boundary velocities at the solid wall. Detailed results will be presented.

## 2 FORMULATION AND NUMERICAL PROCEDURE

Different studies found in the literature concerning flows in triangular cavities use different triangle geometries. For the purpose of being able to compare our results with all these studies, we have considered a general triangle geometry. A general triangle with corner points $\overline{A}$, $\overline{B}$ and $\overline{C}$ where $\overline{A}$ and $\overline{B}$ have the same $y$ position, as shown in Figure 1, can be mapped onto a computational domain, also as shown in Figure 1, using the following relation.

$$\xi = \frac{1}{b_x - a_x} x + \frac{a_x - c_x}{(b_x - a_x)(c_y - h)} y + \frac{c_x h - a_x c_y}{(b_x - a_x)(c_y - h)}$$

$$\eta = \frac{1}{c_y - h} y - \frac{h}{c_y - h} \tag{1}$$

Also the inverse transformation can be done using the following relation

$$y = (c_y - h)\eta + h$$

$$x = (b_x - a_x)\xi - (a_x - c_x)\eta + a_x \tag{2}$$

From these relations, we can calculate the transformation metrics as the following

$$\frac{\partial \xi}{\partial x} = \frac{1}{b_x - a_x} \;,\; \frac{\partial \xi}{\partial y} = \frac{a_x - c_x}{(b_x - a_x)(c_y - h)}$$

$$\frac{\partial \eta}{\partial x} = 0 \;,\; \frac{\partial \eta}{\partial y} = \frac{1}{c_y - h} \tag{3}$$

and also

$$\frac{\partial^2 \xi}{\partial x^2} = \frac{\partial^2 \xi}{\partial y^2} = \frac{\partial^2 \eta}{\partial x^2} = \frac{\partial^2 \eta}{\partial y^2} = 0 \tag{4}$$

The 2-D steady incompressible flow inside a triangular cavity is governed by the Navier-Stokes equations. We consider the N-S equations in streamfunction and vorticity formulation, such that

$$\frac{\partial^2 \psi}{\partial x^2} + \frac{\partial^2 \psi}{\partial y^2} = -\omega \tag{5}$$

$$\frac{1}{Re}\left(\frac{\partial^2 \omega}{\partial x^2} + \frac{\partial^2 \omega}{\partial y^2}\right) = \frac{\partial \psi}{\partial y}\frac{\partial \omega}{\partial x} - \frac{\partial \psi}{\partial x}\frac{\partial \omega}{\partial y} \tag{6}$$

where $\psi$ is streamfunction and $\omega$ is vorticity, $x$ and $y$ are the Cartesian coordinates and $Re$ is the Reynolds number. We note that these equations are non dimensional and a length scale of $(h - c_y)/3$ and a velocity scale of $U$, ie. the velocity of the lid, are used to non-dimensionalize the parameters and the Reynolds number is defined accordingly.

Using the chain rule, the governing N-S equations in general curvilinear coordinates are the following

$$\left(\left(\frac{\partial \xi}{\partial x}\right)^2 + \left(\frac{\partial \xi}{\partial y}\right)^2\right)\frac{\partial^2 \psi}{\partial \xi^2} + \left(\left(\frac{\partial \eta}{\partial x}\right)^2 + \left(\frac{\partial \eta}{\partial y}\right)^2\right)\frac{\partial^2 \psi}{\partial \eta^2}$$

$$+ \left(\frac{\partial^2 \xi}{\partial x^2} + \frac{\partial^2 \xi}{\partial y^2}\right)\frac{\partial \psi}{\partial \xi} + \left(\frac{\partial^2 \eta}{\partial x^2} + \frac{\partial^2 \eta}{\partial y^2}\right)\frac{\partial \psi}{\partial \eta}$$

$$+ 2\left(\frac{\partial \xi}{\partial x}\frac{\partial \eta}{\partial x} + \frac{\partial \xi}{\partial y}\frac{\partial \eta}{\partial y}\right)\frac{\partial^2 \psi}{\partial \xi \partial \eta} = -\omega \tag{7}$$

$$\frac{1}{Re}\Bigg(\left(\left(\frac{\partial \xi}{\partial x}\right)^2 + \left(\frac{\partial \xi}{\partial y}\right)^2\right)\frac{\partial^2 \omega}{\partial \xi^2} + \left(\left(\frac{\partial \eta}{\partial x}\right)^2 + \left(\frac{\partial \eta}{\partial y}\right)^2\right)\frac{\partial^2 \omega}{\partial \eta^2}$$

$$+ \left(\frac{\partial^2 \xi}{\partial x^2} + \frac{\partial^2 \xi}{\partial y^2}\right)\frac{\partial \omega}{\partial \xi} + \left(\frac{\partial^2 \eta}{\partial x^2} + \frac{\partial^2 \eta}{\partial y^2}\right)\frac{\partial \omega}{\partial \eta}$$

$$+ 2\left(\frac{\partial \xi}{\partial x}\frac{\partial \eta}{\partial x} + \frac{\partial \xi}{\partial y}\frac{\partial \eta}{\partial y}\right)\frac{\partial^2 \omega}{\partial \xi \partial \eta}\Bigg)$$

$$= \left(\frac{\partial \xi}{\partial x}\frac{\partial \eta}{\partial y}\right)\frac{\partial \psi}{\partial \eta}\frac{\partial \omega}{\partial \xi} - \left(\frac{\partial \xi}{\partial x}\frac{\partial \eta}{\partial y}\right)\frac{\partial \psi}{\partial \xi}\frac{\partial \omega}{\partial \eta} \tag{8}$$

Substituting for the transformation metrics defined in Equations (3 and 4), we obtain the equations that govern the flow in a triangular cavity shown in Figure 1 as the following

$$A\frac{\partial^2 \psi}{\partial \xi^2} + B\frac{\partial^2 \psi}{\partial \eta^2} + C\frac{\partial^2 \psi}{\partial \xi \partial \eta} = -\omega \tag{9}$$

$$\frac{1}{Re}\left(A\frac{\partial^2 \omega}{\partial \xi^2} + B\frac{\partial^2 \omega}{\partial \eta^2} + C\frac{\partial^2 \omega}{\partial \xi \partial \eta}\right) = D\frac{\partial \psi}{\partial \eta}\frac{\partial \omega}{\partial \xi} - D\frac{\partial \psi}{\partial \xi}\frac{\partial \omega}{\partial \eta} \tag{10}$$

where

$$A = \left(\frac{\partial \xi}{\partial x}\right)^2 + \left(\frac{\partial \xi}{\partial y}\right)^2 \;,\; B = \left(\frac{\partial \eta}{\partial y}\right)^2$$

$$C = 2\frac{\partial \xi}{\partial y}\frac{\partial \eta}{\partial y} \;,\; D = \frac{\partial \xi}{\partial x}\frac{\partial \eta}{\partial y} \tag{11}$$

The boundary conditions in $x$- and $y$-plane are the following

$$\psi = 0 \quad \text{on three sides}$$

$$(u,v) \cdot t = \begin{cases} 1 \text{ on the top side} \\ 0 \text{ on the other two sides} \end{cases}$$

$$(u,v) \cdot n = 0 \quad \text{on three sides} \tag{12}$$

where $n$ is the unit normal vector and $t$ is the tangent vector in clockwise direction and $(u,v)$ are the velocity components in $x$- and $y$-direction where

$$u = \frac{\partial \psi}{\partial y} = \frac{\partial \xi}{\partial y}\frac{\partial \psi}{\partial \xi} + \frac{\partial \eta}{\partial y}\frac{\partial \psi}{\partial \eta}$$
$$v = -\frac{\partial \psi}{\partial x} = -\frac{\partial \xi}{\partial x}\frac{\partial \psi}{\partial \xi} \tag{13}$$

Therefore, in the computational domain, the boundary conditions on the side A-B are

$$\frac{\partial \psi}{\partial \xi} = 0 \quad , \quad \frac{\partial \psi}{\partial \eta} = \frac{\partial y}{\partial \eta} \tag{14}$$

On sides A-C and B-C we have

$$\frac{\partial \psi}{\partial \xi} = 0 \quad , \quad \frac{\partial \psi}{\partial \eta} = 0 \tag{15}$$

Also note that, on side B-C one can also show that the derivative of the streamfunction in the wall normal direction would be homogenous.

Erturk et al. [6] stated that for square driven cavity when fine grids are used, it is possible to obtain numerical solutions at high Reynolds numbers. At high Reynolds numbers thin boundary layers are developed along the solid walls and it becomes essential to use fine grid meshes. Also when fine grids are used, the cell Reynolds number or so called the Peclet number, defined as $Re_c = u\Delta h/\nu$, decreases and this improves the numerical stability (see Weinan and Jian-Guo [22] and Tennehill et al. [20]). In this study, we used a fine grid mesh with $(512 \times 512)/2$ grid points. On this mesh, we solved the governing equations (10 and 11) using SOR method. For the vorticity values at the boundary points we used Thom's formula [21]. We note an important fact that, it is well understood (see Weinan and Jian-Guo [22], Spotz [19], Huang and Wetton [10], Napolitano et al. [16]) that, even though Thom's method is locally first order accurate, the global solution obtained using Thom's method preserve second order accuracy $\mathcal{O}(\Delta h^2)$. On side A-B the vorticity is equal to

$$\omega_{i,0} = -B\frac{2\psi_{i,1}}{\Delta h^2} + B\frac{2}{\Delta h}\frac{\partial y}{\partial \eta} \tag{16}$$

where the subscripts $i$ and $j$ are the grid indices in $\xi$- and $\eta$-directions respectively and the grid index 0 refers to points on the wall and 1 refers to points adjacent to the wall and also $\Delta h$ refers to grid spacing.

Similarly on side A-C the vorticity is equal to

$$\omega_{0,j} = -A\frac{2\psi_{1,j}}{\Delta h^2} \tag{17}$$

and on side B-C, it is equal to

$$\omega_{i,N-i} = -A\frac{2\psi_{i-1,N-i}}{\Delta h^2} - B\frac{2\psi_{i,N-i-1}}{\Delta h^2} - C\frac{2\psi_{i-1,N-i-1}}{4\Delta h^2} \tag{18}$$

## 3 RESULTS AND DISCUSSIONS

During the iterations, as a measure of convergence, we monitored several residuals. We define the residual of the steady state streamfunction and vorticity Equations (10 and 11), as the following

$$R_\psi = A\frac{\psi_{i-1,j} - 2\psi_{i,j} + \psi_{i+1,j}}{\Delta h^2} + B\frac{\psi_{i,j-1} - 2\psi_{i,j} + \psi_{i,j+1}}{\Delta h^2}$$
$$+ C\frac{\psi_{i+1,j+1} + \psi_{i-1,j-1} - \psi_{i+1,j-1} - \psi_{i-1,j+1}}{4\Delta h^2} + \omega_{i,j} \tag{19}$$

$$R_\omega = \frac{A}{Re}\frac{\omega_{i-1,j} - 2\omega_{i,j} + \omega_{i+1,j}}{\Delta h^2} + \frac{B}{Re}\frac{\omega_{i,j-1} - 2\omega_{i,j} + \omega_{i,j+1}}{\Delta h^2}$$
$$+ \frac{C}{Re}\frac{\omega_{i+1,j+1} + \omega_{i-1,j-1} - \omega_{i+1,j-1} - \omega_{i-1,j+1}}{4\Delta h^2}$$
$$- D\frac{\psi_{i,j+1} - \psi_{i,j-1}}{2\Delta h}\frac{\omega_{i+1,j} - \omega_{i-1,j}}{2\Delta h}$$
$$+ D\frac{\psi_{i+1,j} - \psi_{i-1,j}}{2\Delta h}\frac{\omega_{i,j+1} - \omega_{i,j-1}}{2\Delta h} \tag{20}$$

These residuals indicate the degree to which the numerical solution has converged to steady state. In our computations, in all of the triangle geometries considered, at any Reynolds numbers, we considered that the convergence was achieved when the absolute values of the maximum of the residuals defined above in the computational domain $\left(\max(|R_\psi|) \text{ and } \max(|R_\omega|)\right)$ were both below $10^{-10}$. Such a low convergence level is chosen to ensure the accuracy of the solution. At these convergence levels, the maximum absolute change in the streamfunction variable in the computational domain $\left(\max(|\psi_{i,j}^{n+1} - \psi_{i,j}^n|)\right)$ was in the order of $10^{-16}$ and for the vorticity variable $\left(\max(|\omega_{i,j}^{n+1} - \omega_{i,j}^n|)\right)$ it was in the order of $10^{-14}$. In other words, our numerical solutions of the streamfunction and the vorticity variables are accurate up to 15 and 13 digits respectively. Also at these convergence levels, the maximum absolute normalized change in the streamfunction and vorticity variable in the computational domain $\left(\max(|(\psi_{i,j}^{n+1} - \psi_{i,j}^n)/\psi_{i,j}^n|) \text{ and } \max(|(\omega_{i,j}^{n+1} - \omega_{i,j}^n)/\omega_{i,j}^n|)\right)$ were in the order of $10^{-13}$ and $10^{-12}$ respectively. In other words, at convergence the streamfunction variable was changing with $10^{-11}$ percent of its value and the vorticity variable was changing $10^{-10}$ percent of its value, at each iteration step. These numbers ensure that our numerical solutions are indeed very accurate.

In this study we considered different triangle geometries. In order to demonstrate how we define the Reynolds number, let us consider a dimensional equilateral triangle with coordinates of corner points

$$a_x = -\sqrt{3}a \, , \, b_x = \sqrt{3}a \, , \, h = a \, , \, c_x = 0 \, , \, c_y = -2a \quad (21)$$

If we use the top wall velocity $U$ and the length scale $a$ in nondimensionalization, then the Reynolds number is defined as $Ua/\nu$ and the coordinates of the corner points of the nondimensional equilateral triangle become

$$a_x = -\sqrt{3} \, , \, b_x = \sqrt{3} \, , \, h = 1 \, , \, c_x = 0 \, , \, c_y = -2 \quad (22)$$

We, first considered this equilateral triangle geometry which was also considered by McQuain et al. [14], Ribbens et al. [17], Jyotsna and Vanka [11], Li and Tang [13] and Gaskell et al. [8]. We solved the flow in this triangle cavity at various Reynolds numbers ranging between 1 and 1750. We note that, if the length of one side of the triangle ($2\sqrt{3}a$) was used in nondimensionalization, as it was used by Gaskell et al. [8], then our Reynolds number of 1750 would be $2\sqrt{3}$ fold such that it would correspond to a Reynolds number of 6062. Qualitatively, Figure 2 shows the streamline contours of the flow at various Reynolds numbers and also Figure 3 shows the vorticity contours at the same Reynolds numbers. In terms of quantitative analysis, Table 1 tabulates the center of the primary eddy and the streamfunction and vorticity values at the core, together with results found in the literature. Our results agree well with McQuain et al. [14], Ribbens et al. [17] and agree excellent with Gaskell et al. [8] up to the maximum Reynolds number ($Re$=500) they have considered. However, the results of Li and Tang [13] differ from our results. The difference can be seen more obvious in Figure 4, where we plot the vorticity values at the center of the primary vortex tabulated in Table 1, with respect to the Reynolds number. The results of Li and Tang [13] start to behave differently starting from $Re$=100 from the rest of the results. Also in Figure 4, the results of McQuain et al. [14] and Ribbens et al. [17] start to deviate from present results and Gaskell et al.'s [8] results after $Re$=200. Having a different behavior, the vorticity value of McQuain et al. [14] and Ribbens et al. [17] shows an increase such that their vorticity value at $Re$=500 is greater than the vorticity value at $Re$=350, whereas the present results and Gaskell et al.'s [8] results show a continuous decrease. We believe that these behaviors are due to the coarse grids used in those studies and in order to resolve these behaviors we decided to solve the same flow problem using several coarse grid meshes. We computed the flow using (64×64)/2, (128×128)/2 and (256×256)/2 grid points also. For a given grid mesh we have solved the flow for increasing Reynolds number until a particular $Re$ where the solution was not convergent but oscillating. For this particular $Re$ when the number of grid points was increased, the convergence was recovered and we were able to obtain a solution.

In order to demonstrate the effect of the grid mesh on the numerical solution, in Figure 5 we plotted the vorticity value at the center of the primary vortex obtained using different grid mesh, with respect to Reynolds number. From the results in Figure 5, we conclude that (64×64)/2 grid mesh is too coarse for this flow problem. In this figure, looking at the solution of (128×128)/2 and (256×256)/2, the vorticity value decreases with increasing Reynolds number however shows an increase at the largest Reynolds number where beyond that $Re$ the solution is not converging. We note that this behavior is similar to that of McQuain et al. [14] and Ribbens et al. [17] in Figure 4.

In Figure 5 looking at the solutions of different grid mesh, for example at $Re$=350, we see that as the number of grid points in the mesh is increased, the vorticity value at the center of the primary eddy is decreased however this decrease slows down as $\Delta h$ gets smaller. This is not surprising since the solution gets more accurate as the number of grid points is increased, and also since our solution is spatially second order accurate, $\mathcal{O}(\Delta h^2)$, the computed vorticity value should be expected to converge to the real value with $\Delta h^2$. At the finest grid mesh, (512×512)/2, at high Reynolds numbers our the vorticity value seems to stagnate around 1.16. We note that, we expect the vorticity value to be a little lower than this value, if the number of grid points is increased furthermore. However increasing the number of grid number furthermore increases the computing time, therefore we will not consider a finer grid mesh than (512×512)/2. This grid study shows that fine grids are necessary for accurate solutions for triangular cavity flow especially at high Reynolds numbers.

McQuain et al. [14] analytically obtained the constant vorticity value of the inviscid rotational flow at infinite Reynolds number as $\sqrt{10}/3$=1.054 using Batchelor's mean-square law. This infinite Reynolds number value of the vorticity of 1.054 is shown with the dotted line in Figure 4. Our computed primary eddy vorticity asymptotes to a value around 1.16 as the Reynolds number increases, which is a little higher than the theoretical value. We note that the theory, assumes that at infinite Reynolds number the whole inviscid fluid rotates as a solid body with a constant vorticity which is coupled to boundary layer flows at the solid surface. However in reality, from Figure 2 it is evident that the whole fluid does not rotate as a solid body, instead there appears progressively smaller counter-rotating recirculating regions at the bottom corner and also one towards the upper left corner. We note that, especially the eddies at the bottom corner occupy some portion of the corner as the Reynolds number increases. However, the increase in the size of the portion of the bottom corner eddies almost stops after $Re$=500 and the size of the primary eddy remains almost constant beyond and so is the value of the vorticity at the core of the primary eddy. It looks like, for an equilateral triangular cavity flow at high Reynolds numbers, the mean square law predicts the strength of the primary eddy within an error due to the effect of the secondary eddies. For circular or elliptic boundaries Ribbens et al. [17], for square cavity flow Erturk *et al.* [6], and for rectangle cavities McQuain *et al.* [14] have shown that the mean square law is approximately valid and successful in predicting the strength of the primary eddy at high Reynolds numbers. Due to small corner angles in a triangle geometry, for triangular cavities the mean square law is not as successful as it was in other geometries.

We then considered an isosceles triangle which was also considered by Jyotsna and Vanka [11], where

$$a_x = -1, \; b_x = 1, \; h = 0, \; c_x = 0, \; c_y = -4 \qquad (23)$$

We note that, our definition of Reynolds number is equivalent to one fourth of the Reynolds number definition used by Jyotsna and Vanka [11].

Figure 6 shows the streamline contours of the flow in this triangle at various Reynolds numbers. Table 2 compares the location of the primary eddy center with that of Jyotsna and Vanka [11] and Gaskell *et al.* [8]. The results agree with each other, although we believe that our results are more accurate.

Moffatt [15] analytically studied the recirculating eddies near a sharp corner in the Stokes regime and predicted that the distance from the corner to the centers of eddies and also the velocity at the dividing streamline between the eddies, follow a geometric sequence.

Using the considered triangle geometry, Jyotsna and Vanka [11] have used the small Reynolds number solutions to compare with Moffatt's [15] Stokes regime predictions. However, when Reynolds number increases the asymmetry in the flow field increases and the comparison with the Stokes regime could not be possible. In order to be able to compare our results with Moffatt [15], we solve the considered triangle geometry for the Stokes regime, i.e. $Re=0$. The solution at $Re=0$, as seen in Figure 6a, is symmetrical. For this Stokes regime solution, we first calculate the $u$-velocity along a perpendicular line from the bottom corner to the top wall. In this velocity profile, the points where the $u$-velocity is zero correspond to the eddy centers. The distance of these points to the bottom corner would give us $r_n$, where we use in calculating the ratio $r_n/r_{n+1}$. In the velocity profile, the points where the $u$-velocity has local maximum correspond to the $u$-velocity at the dividing streamline between the eddies, where Moffatt [15] have used these velocities as a measure of the intensity of consecutive eddies. These local maximum $u$-velocities would give us $I_n$, where we use in calculating the ratio $I_n/I_{n+1}$. For the considered triangle geometry where $\theta=28.072°$, Table 3 tabulates the calculated ratios of $r_n/r_{n+1}$ and $I_n/I_{n+1}$ along with analytical predictions of Moffatt [15] and the agreement is good.

We then considered an isosceles triangle which was also considered by Gaskell *et al.* [8], with

$$a_x = -1, \; b_x = 1, \; h = 0, \; c_x = 0, \; c_y = \frac{-1}{\tan(\frac{\theta}{2})} \qquad (24)$$

where $\theta=20°, 30°, 40°, 50°, 60°, 70°, 80°$ and $90°$. We will consider Stokes flow in these triangles, such that we solve for $Re=0$. The analytical predictions of Moffatt [15] for these $\theta$ values, are tabulated in Table 4. Figure 7 qualitatively shows the change in the Stokes flow in an isosceles triangle as the corner angle, $\theta$, changes. For this case, our computed relative positions and intensities of eddies are tabulated in Table 5 along with results of Gaskell *et al.* [8]. The agreement between our computed results and that of Gaskell *et al.* [8] and also predictions of Moffatt [15] is good. We note that, Moffatt's [15] analytical analysis predicts that there will be infinite sequence of eddies at the bottom corner. This result is, of course, impossible to capture for a numerical analysis due to the resolution of the grid mesh. However since we have used a finer grid mesh, with $(512\times512)/2$ grid points, than the grid mesh used by Gaskell *et al.* [8], with 6605 nodes, we were able to capture more sequence of eddies than Gaskell *et al.* [8], at the sharp bottom stagnant corner.

We then considered an isosceles right triangle with the 90° corner being at the top right corner, such that with corner points

$$a_x = 0, \; b_x = 1, \; h = 1, \; c_x = 1, \; c_y = 0 \qquad (25)$$

The flow in this considered triangle is solved up to $Re=7500$. Figure 8 shows the flow topology as the Reynolds number increases and Table 6 tabulates the properties of the primary eddy, the streamfunction and the vorticity value at the center of the eddy and also the location of the center, for future references.

We then also considered an isosceles right triangle with the 90° corner being at the top left corner, such that with corner points

$$a_x = 0, \; b_x = 1, \; h = 1, \; c_x = 0, \; c_y = 0 \qquad (26)$$

Using this triangle geometry, we were able to obtain solutions up to $Re=2500$. Figure 9 shows the flow topology as a function of Reynolds numbers. In Table 7, for the considered triangle geometry, we tabulated the properties of the primary eddy, i.e. the eddy that is closest to the moving lid, for future references. As it is obvious from Figures 8 and 9, in both cases the flow behaves very differently as the Reynolds number increases, which shows that the flow structures in a triangle cavity are greatly affected by the triangle geometry.

## 4 CONCLUSIONS

In this study we have presented high accurate, fine grid solutions of 2-D steady incompressible flow in triangle cavities. The governing equations are solved up to very low residuals at various Reynolds numbers. The computed solutions are compared with numerical solutions of McQuain *et al.* [14], Ribbens *et al.* [17], Jyotsna and Vanka [11], Li and Tang [13] and Gaskell *et al.* [8] and also with analytical predictions of Moffatt [15], and good agreement is found. Our results showed that for an equilateral triangular cavity flow, Batchelor's mean-square law is not as successful as it was in square or rectangle cavity flows, due to small stagnant corner angle.

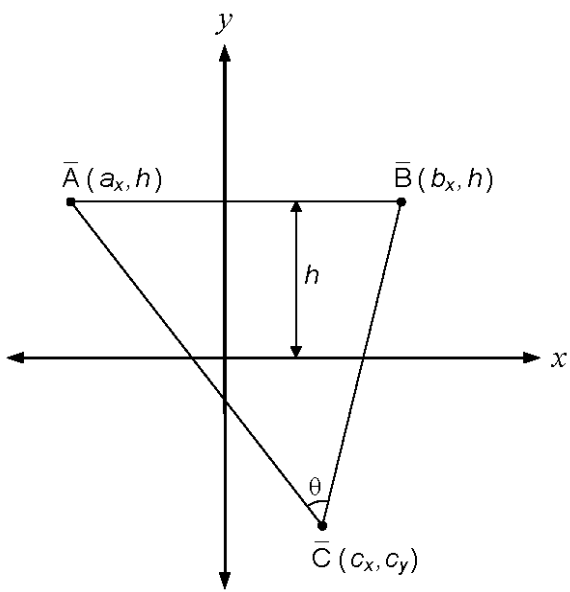 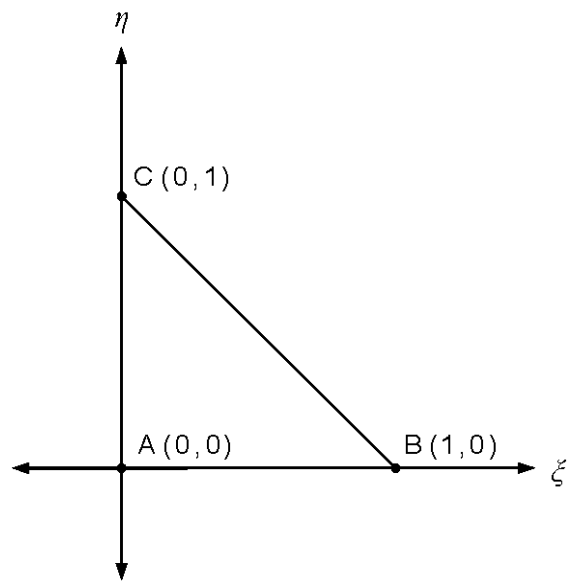

a) Physical Domain  b) Computational Domain

Figure 1)   Geometric transformation of the triangular cavity

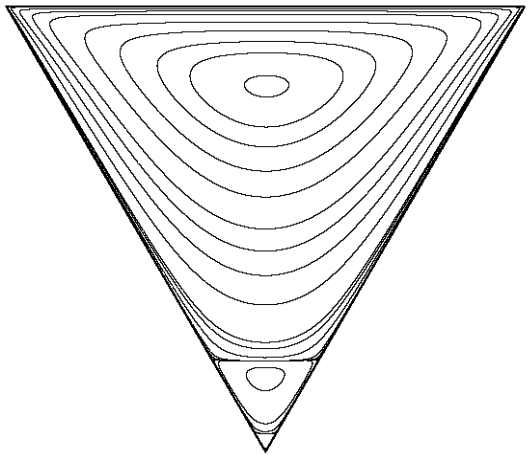

a) Re=1

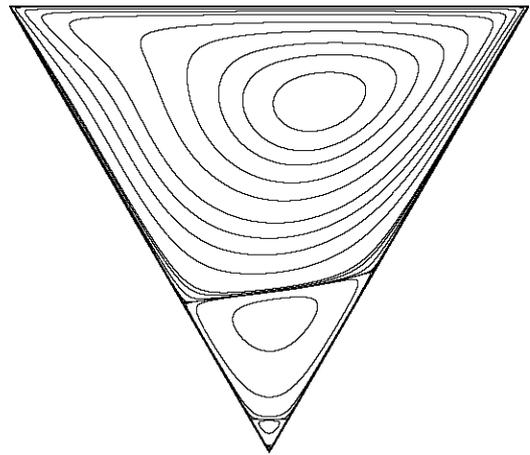

b) Re=100

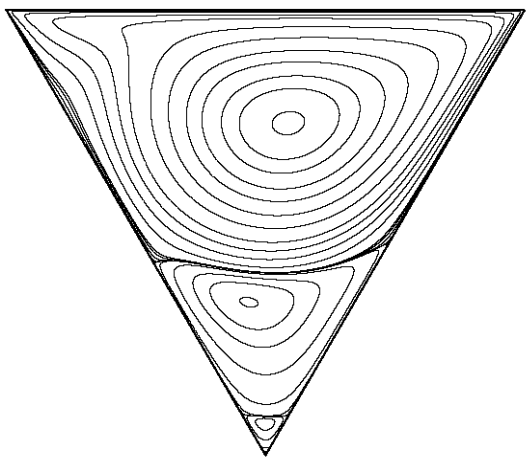

c) Re=350

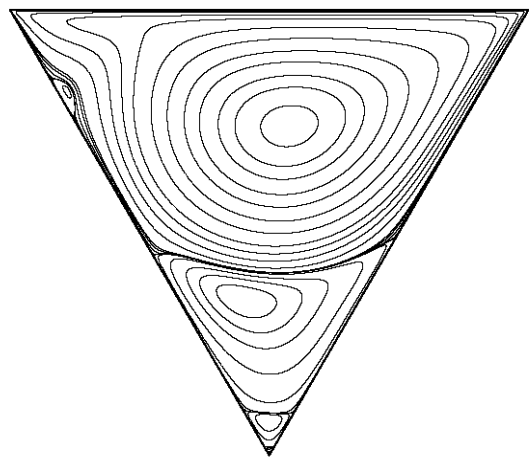

d) Re=500

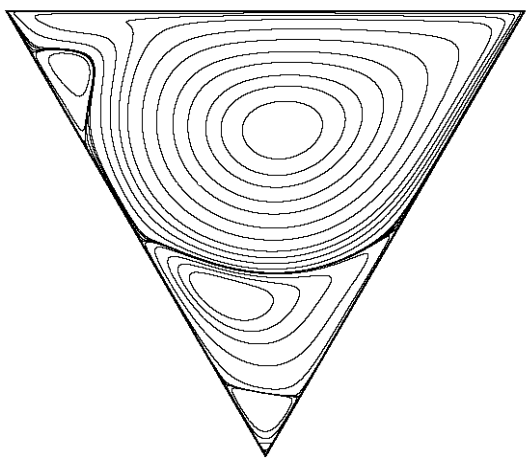

e) Re=1000

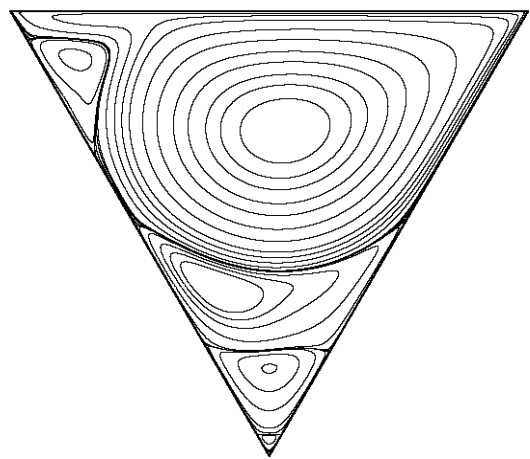

f) Re=1750

Figure 2)   Streamline contours at various Reynolds numbers

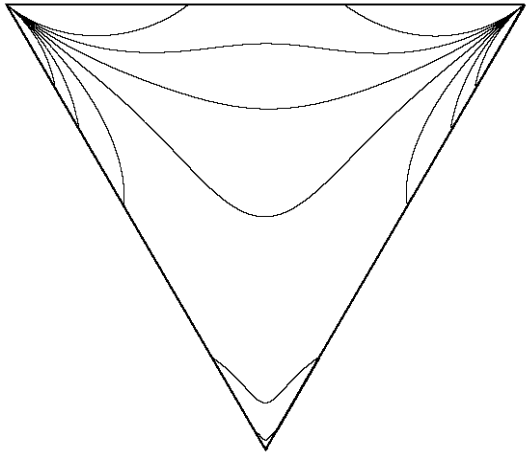
a)  Re=1

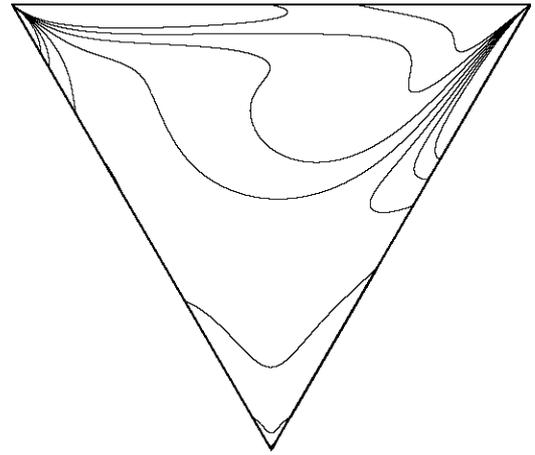
b)  Re=100

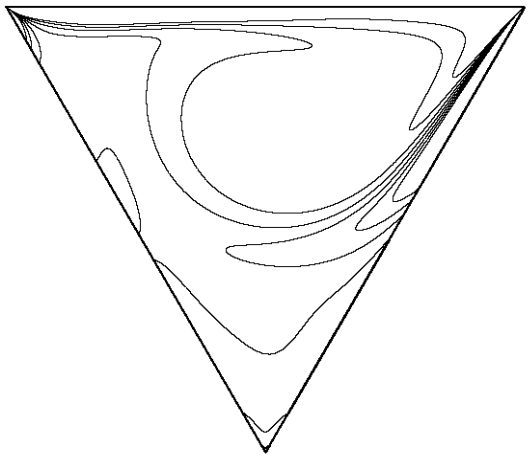
c)  Re=350

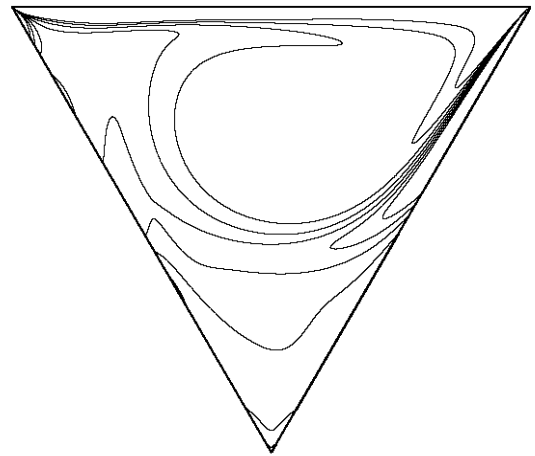
d)  Re=500

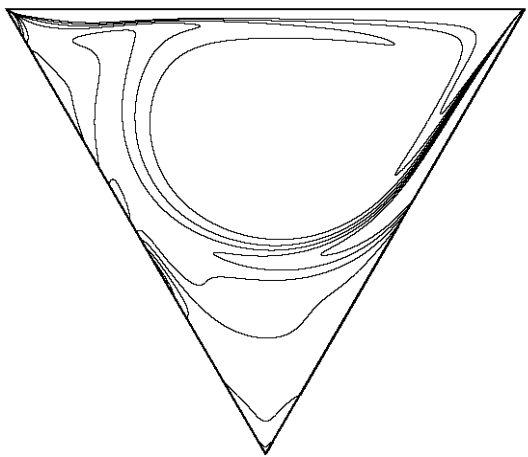
e)  Re=1000

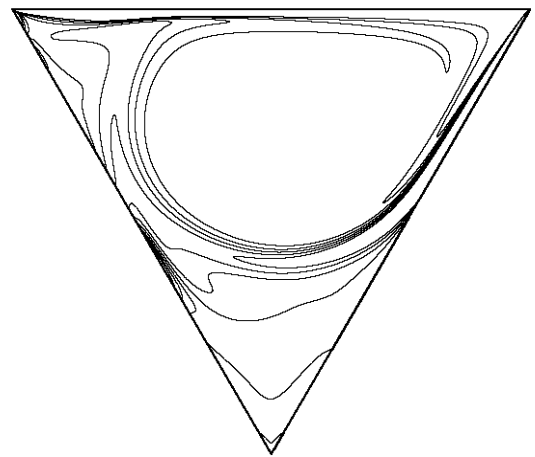
f)  Re=1750

Figure 3)   Vorticity contours at various Reynolds numbers

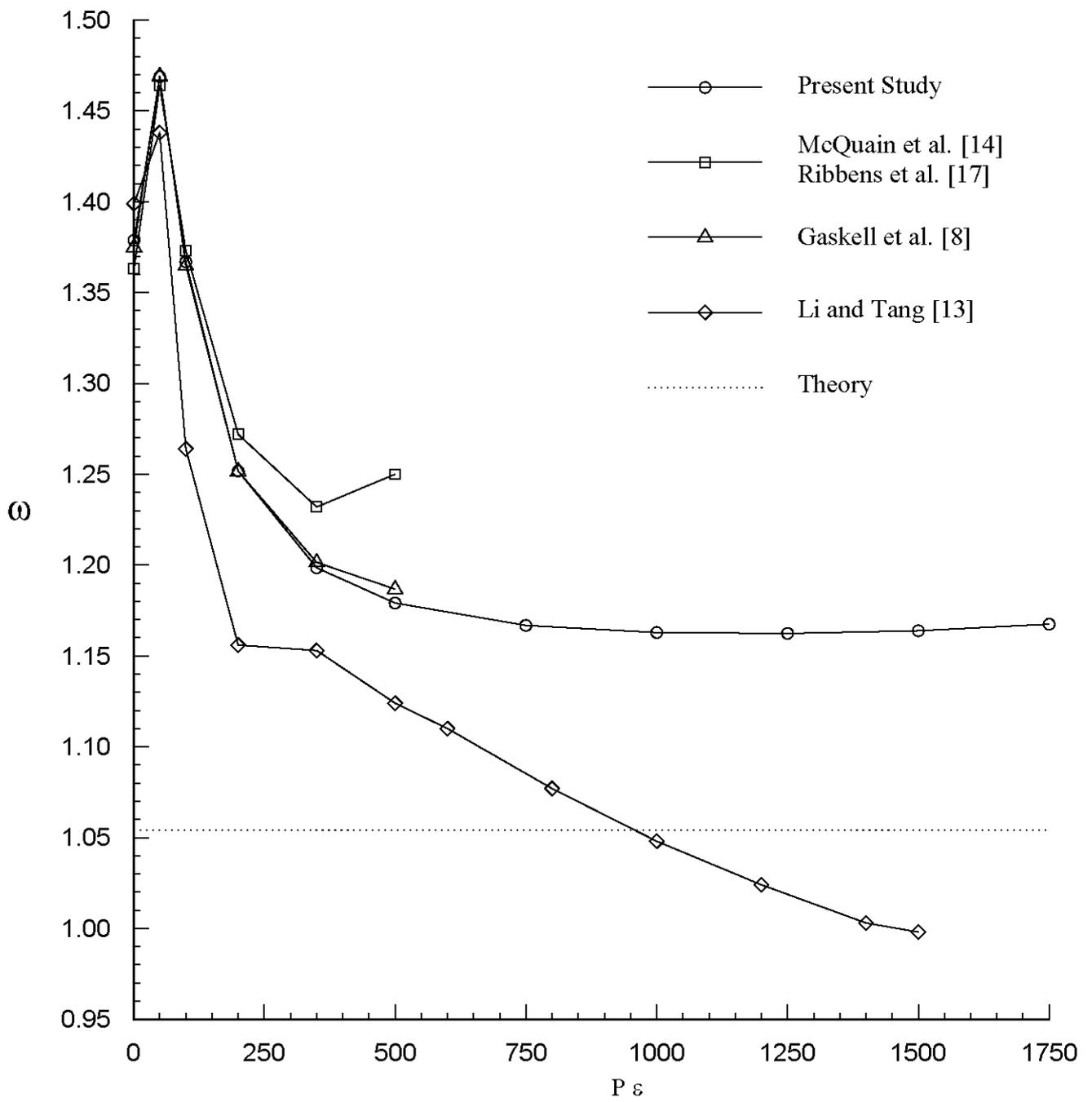

Figure 4) Comparison of the vorticity at the center of the primary eddy

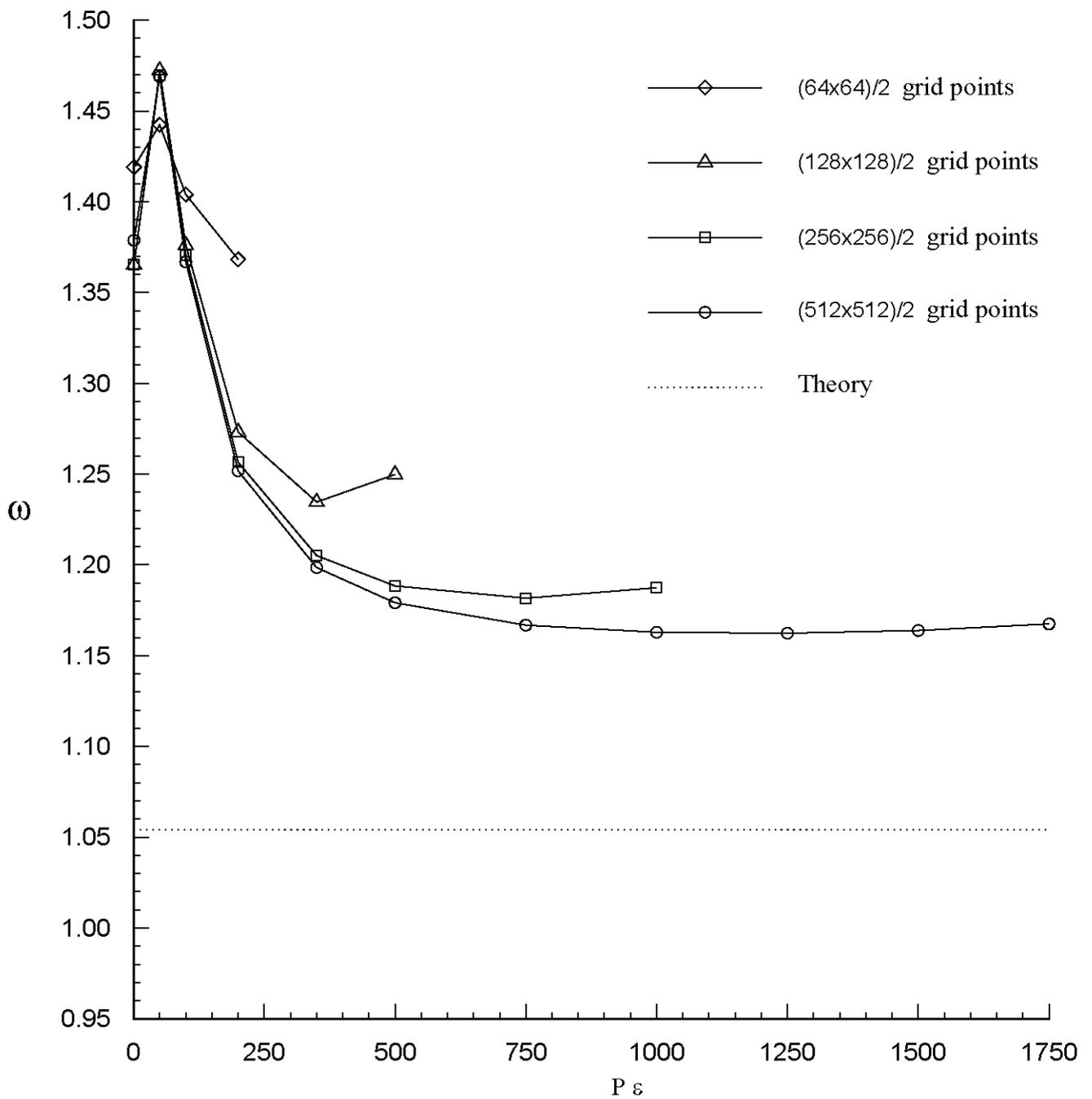

Figure 5) Vorticity at the center of the primary eddy obtained with different grid mesh

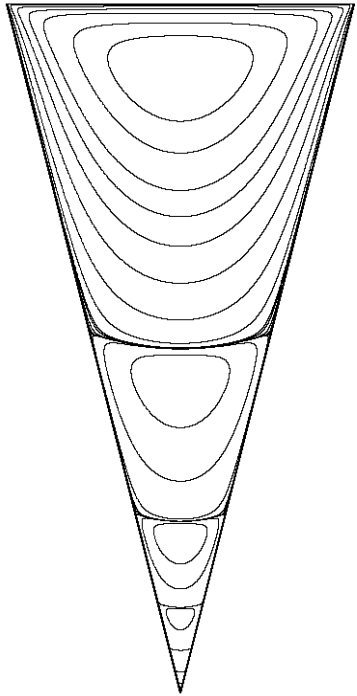 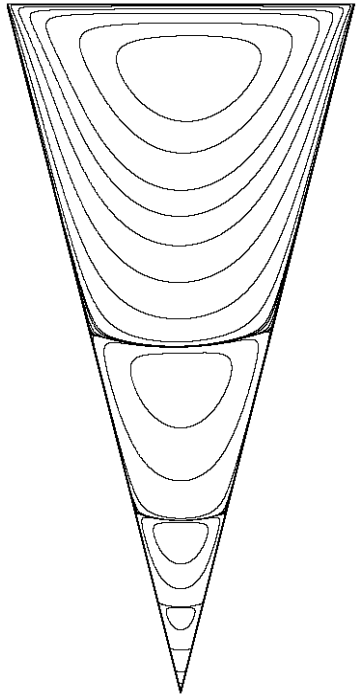 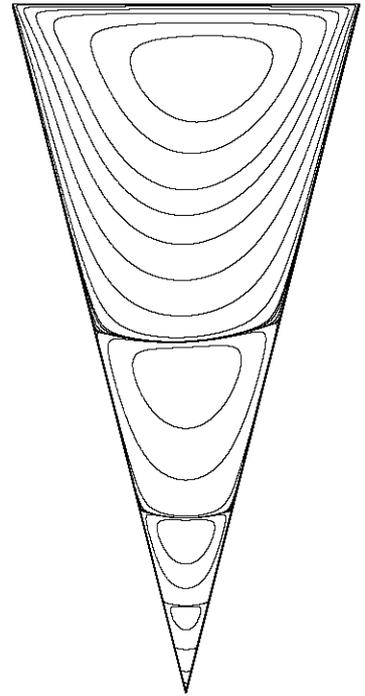

a)  Re=0            b)  Re=12.5          c)  Re=25

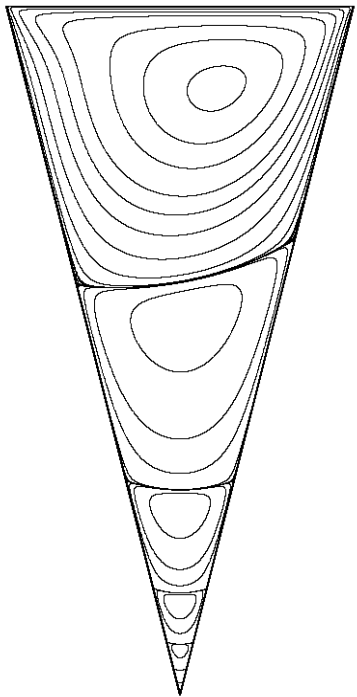 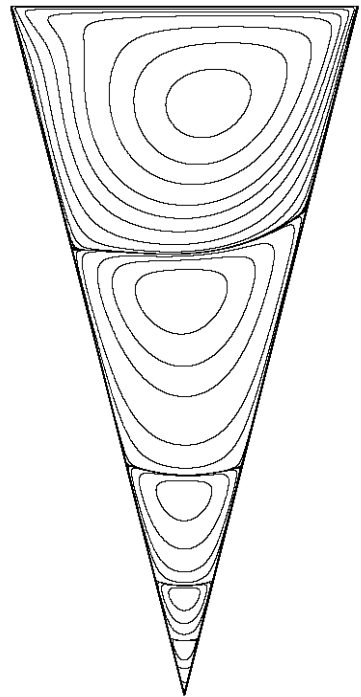

d)  Re=100          e)  Re=200

Figure 6)   Contour figures of isosceles triangle

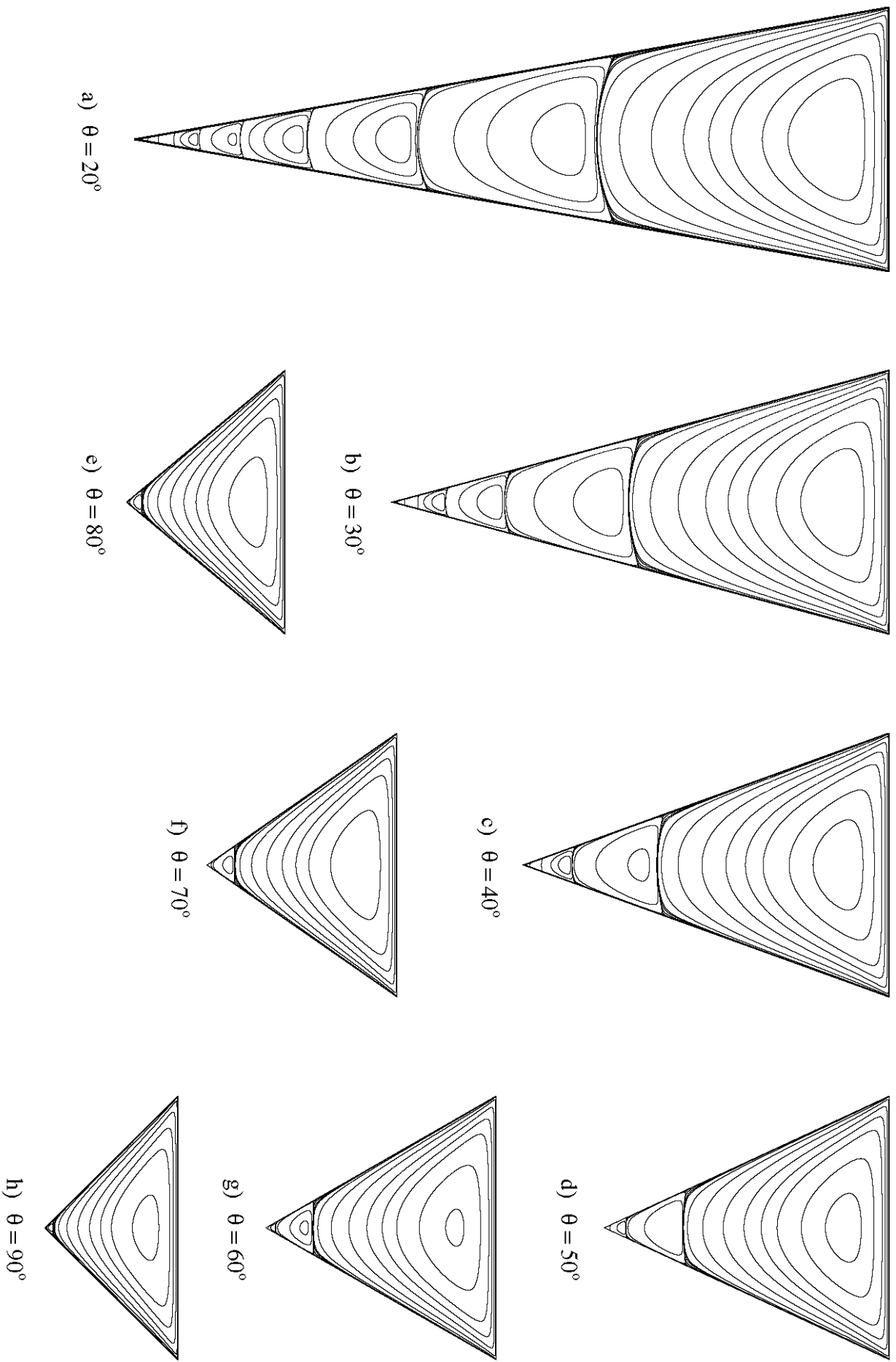

Figure 7) Contour figures of isosceles triangle with various corner angle

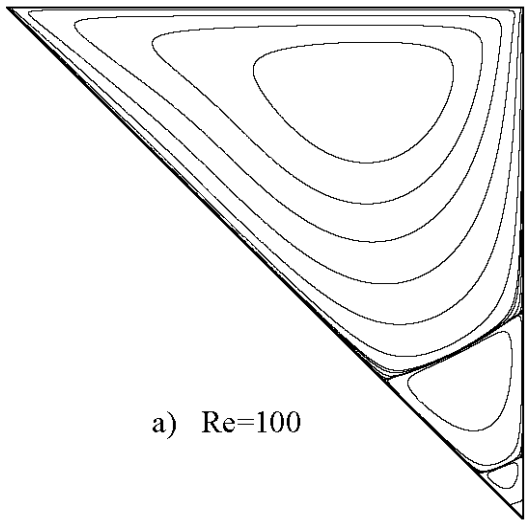 a) Re=100

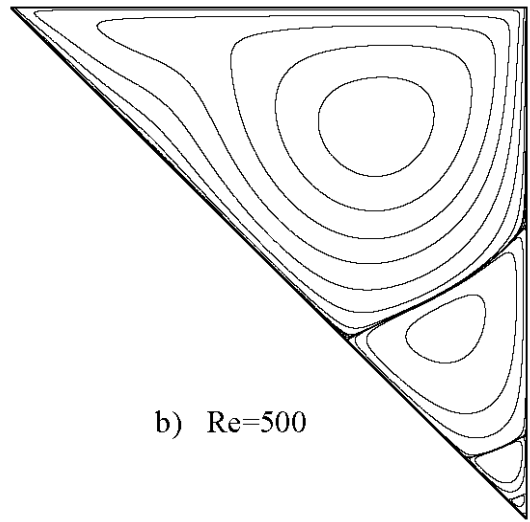 b) Re=500

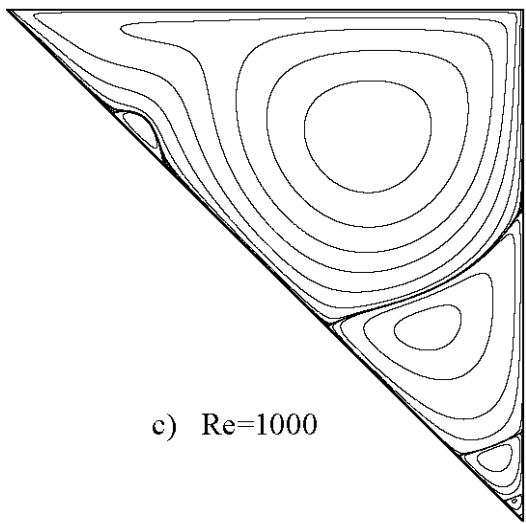 c) Re=1000

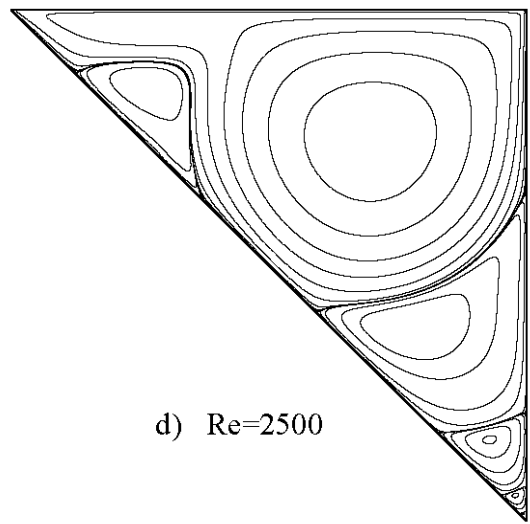 d) Re=2500

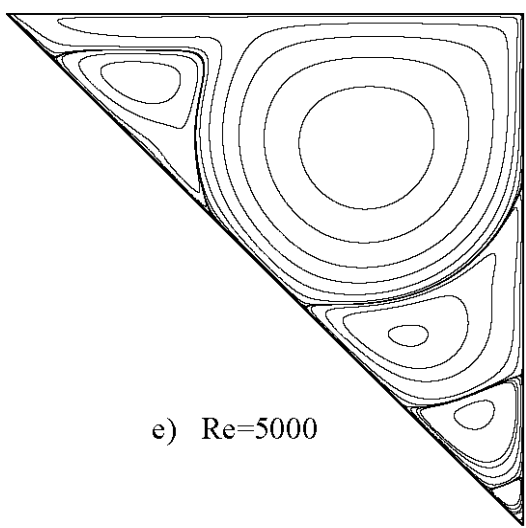 e) Re=5000

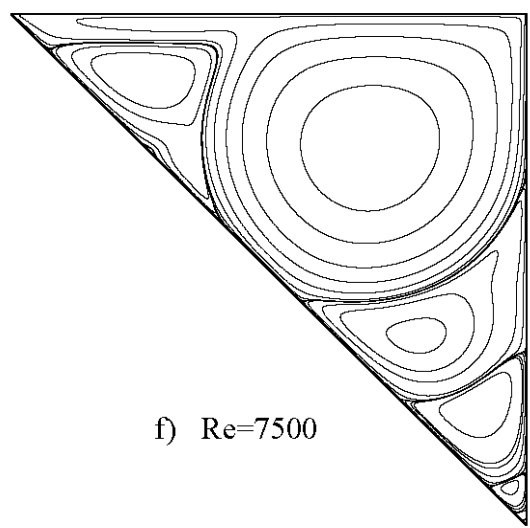 f) Re=7500

Figure 8) Contour figures of right hand side aligned right triangle at various Re.

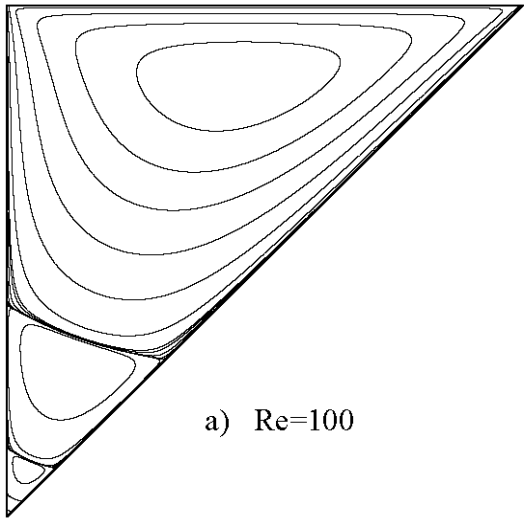
a) Re=100

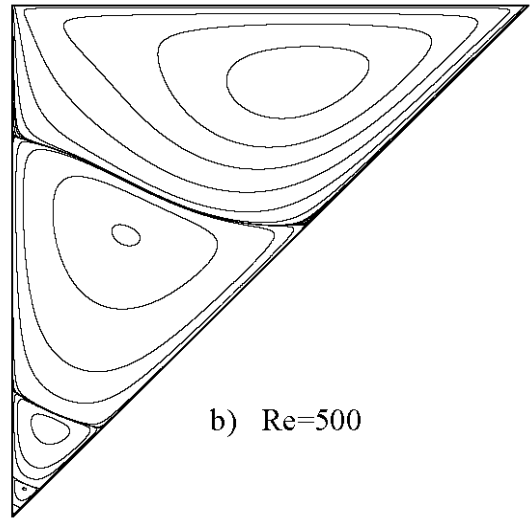
b) Re=500

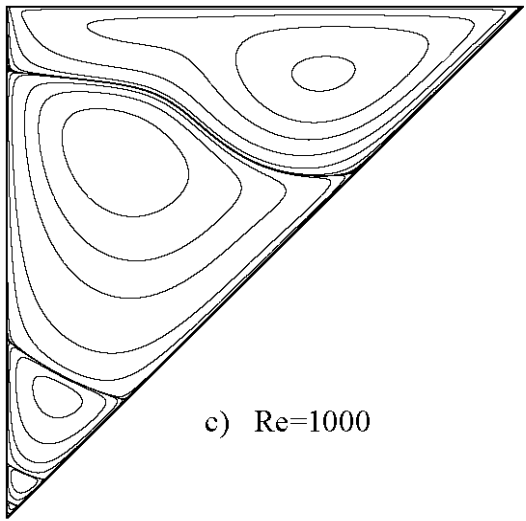
c) Re=1000

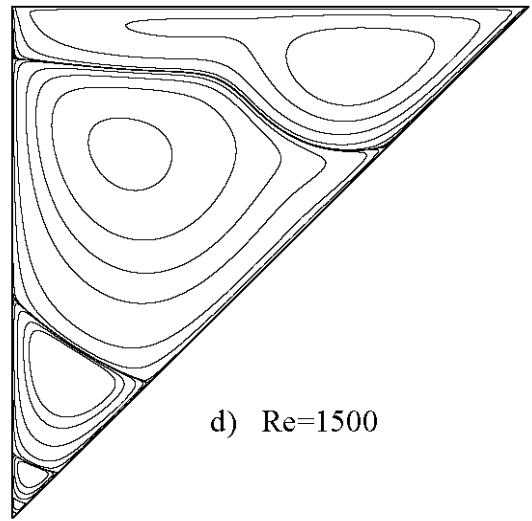
d) Re=1500

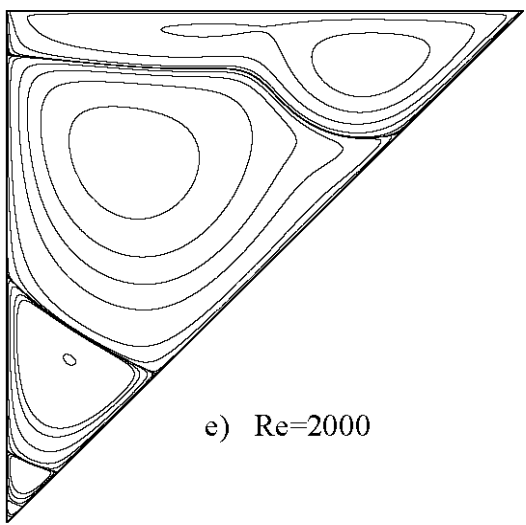
e) Re=2000

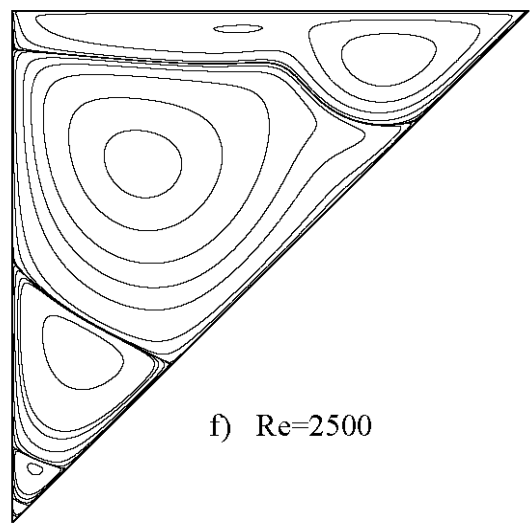
f) Re=2500

Figure 9) Contour figures of left hand side aligned right triangle at various Re.

| | Present Study 512×512 grid mesh | Gaskell *et al.* [8] 6605 nodes | McQuain *et al.* [14] Ribbens *et al.* [17] 200×200 grid mesh | Li and Tang [13] 80×80 grid mesh |
|---|---|---|---|---|
| Re | $\psi \quad \omega$ $(x,y)$ | $\psi \quad \omega$ $(x,y)$ | $\psi \quad \omega$ $(x,y)$ | $\psi \quad \omega$ $(x,y)$ |
| 1 | -0.2329  -1.3788 $(0.0101, 0.4668)$ | -0.232  -1.375 $(0.010, 0.467)$ | -0.233  -1.363 $(0.017, 0.460)$ | -0.234  -1.399 $(0.000, 0.475)$ |
| 50 | -0.2369  -1.4689 $(0.3484, 0.4434)$ | -0.236  -1.469 $(0.350, 0.442)$ | -0.237  -1.464 $(0.346, 0.445)$ | -0.235  -1.438 $(0.368, 0.438)$ |
| 100 | -0.2482  -1.3669 $(0.3315, 0.3555)$ | -0.249  -1.365 $(0.333, 0.356)$ | -0.247  -1.373 $(0.329, 0.355)$ | -0.244  -1.264 $(0.368, 0.363)$ |
| 200 | -0.2624  -1.2518 $(0.2030, 0.2734)$ | -0.263  -1.252 $(0.208, 0.273)$ | -0.260  -1.272 $(0.208, 0.280)$ | -0.262  -1.156 $(0.173, 0.250)$ |
| 350 | -0.2724  -1.1985 $(0.1556, 0.2383)$ | -0.274  -1.201 $(0.149, 0.234)$ | -0.268  -1.232 $(0.173, 0.265)$ | -0.274  -1.153 $(0.152, 0.213)$ |
| 500 | -0.2774  -1.1791 $(0.1319, 0.2207)$ | -0.281  -1.187 $(0.128, 0.217)$ | -0.269  -1.250 $(0.173, 0.265)$ | -0.278  -1.124 $(0.108, 0.213)$ |
| 750 | -0.2818  -1.1668 $(0.1150, 0.2031)$ | - | - | - |
| 1000 | -0.2844  -1.1629 $(0.1116, 0.1973)$ | - | - | -0.279  -1.048 $(0.108, 0.138)$ |
| 1250 | -0.2861  -1.1624 $(0.1049, 0.1973)$ | - | - | - |
| 1500 | -0.2873  -1.1639 $(0.1015, 0.1914)$ | - | - | -0.277  -0.998 $(0.108, 0.138)$ |
| 1750 | -0.2881  -1.1675 $(0.1015, 0.1914)$ | - | - | - |

Table 1)  Comparison of primary eddy properties, for equilateral triangle with corner points $a_x = -\sqrt{3}$, $b_x = \sqrt{3}$, $h = 1$, $c_x = 0$, $c_y = -2$

| Re | Present Study $(x,y)$ | Gaskell et al. [8] $(x,y)$ | Jyotsna and Vanka [11] $(x,y)$ |
|---|---|---|---|
| 12.5 | (0.059,-0.391) | (0.060,-0.392) | (0.050,-0.395) |
| 25 | (0.115,-0.398) | (0.114,-0.396) | (0.097,-0.400) |
| 100 | (0.213,-0.477) | (0.212,-0.478) | (0.225,-0.450) |
| 200 | (0.129,-0.563) | (0.124,-0.568) | (0.153,-0.545) |

Table 2) Comparison of the location of the primary eddy center for isosceles triangle with corner points $a_x = -1$, $b_x = 1$, $h = 0$, $c_x = 0$, $c_y = -4$

| Present Study | | | | | | | | | | Moffatt [15] | |
|---|---|---|---|---|---|---|---|---|---|---|---|
| $r_1/r_2$ | $r_2/r_3$ | $r_3/r_4$ | $r_4/r_5$ | $r_5/r_6$ | $I_1/I_2$ | $I_2/I_3$ | $I_3/I_4$ | $I_4/I_5$ | $I_5/I_6$ | $r_n/r_{n+1}$ | $I_n/I_{n+1}$ |
| 1.99 | 2.01 | 2.01 | 2.03 | 2.09 | 386.0 | 407.9 | 412.8 | 432.8 | 513.3 | 2.01 | 407 |

Table 3) Relative eddy center locations $r_n/r_{n+1}$ and intensities $I_n/I_{n+1}$, for isosceles triangle with $\theta = 28.072°$

| $\theta$ | $r_n/r_{n+1}$ | $I_n/I_{n+1}$ |
|---|---|---|
| 20° | 1.64 | 381.5 |
| 30° | 2.11 | 414.5 |
| 40° | 2.76 | 467.9 |
| 50° | 3.67 | 552.2 |
| 60° | 5.00 | 687.4 |
| 70° | 7.05 | 915.2 |
| 80° | 10.44 | 1331.1 |
| 90° | 16.57 | 2189.1 |

Table 4) Moffatt's [15] predictions of relative eddy center locations $r_n/r_{n+1}$ and intensities $I_n/I_{n+1}$ for Stokes regime for various $\theta$

| | Present Study | | | | | | | | | | | | | Gaskell et al. [8] | | | | | | | |
|---|---|---|---|---|---|---|---|---|---|---|---|---|---|---|---|---|---|---|---|---|---|
| | $r_1/r_2$ | $r_2/r_3$ | $r_3/r_4$ | $r_4/r_5$ | $r_5/r_6$ | $r_6/r_7$ | $I_1/I_2$ | $I_2/I_3$ | $I_3/I_4$ | $I_4/I_5$ | $I_5/I_6$ | $I_6/I_7$ | $r_1/r_2$ | $r_2/r_3$ | $r_3/r_4$ | $I_1/I_2$ | $I_2/I_3$ | $I_3/I_4$ |
| $\theta=20°$ | 1.62 | 1.64 | 1.64 | 1.64 | 1.65 | 1.66 | 361.5 | 382.6 | 386.7 | 386.2 | 404.8 | 462.9 | 1.62 | 1.63 | 1.61 | 362.0 | 384.9 | 394.2 |
| $\theta=30°$ | 2.09 | 2.11 | 2.12 | 2.15 | - | - | 383.4 | 416.8 | 419.9 | 457.3 | - | - | 2.09 | 2.11 | 2.02 | 392.7 | 421.2 | 450.5 |
| $\theta=40°$ | 2.72 | 2.76 | 2.80 | - | - | - | 445.2 | 473.8 | 490.5 | - | - | - | 2.72 | 2.71 | 2.35 | 447.6 | 474.3 | 341.8 |
| $\theta=50°$ | 3.60 | 3.68 | - | - | - | - | 527.0 | 569.9 | - | - | - | - | 3.59 | 3.48 | - | 531.4 | 406.5 | - |
| $\theta=60°$ | 4.89 | 5.04 | - | - | - | - | 657.0 | 713.5 | - | - | - | - | 4.88 | 4.69 | - | 654.2 | 1079.1 | - |
| $\theta=70°$ | 6.86 | 7.20 | - | - | - | - | 876.6 | 1082.1 | - | - | - | - | 6.81 | - | - | 888.0 | - | - |
| $\theta=80°$ | 10.09 | - | - | - | - | - | 1283.1 | - | - | - | - | - | 10.3 | - | - | 1307.8 | - | - |
| $\theta=90°$ | 15.89 | - | - | - | - | - | 2125.5 | - | - | - | - | - | 15.1 | - | - | 2069.8 | - | - |

Table 5) Comparison of relative eddy center locations $r_n/r_{n+1}$ and intensities $I_n/I_{n+1}$, for Stokes regime in isosceles triangle with $\theta=20°$, $30°$, $40°$, $50°$, $60°$, $70°$, $80°$ and $90°$

| Re | $\psi$ | $\omega$ | $(x, y)$ |
|---|---|---|---|
| 100 | -0.07162 | -4.83000 | (0.7090, 0.8320) |
| 500 | -0.08106 | -4.12104 | (0.7070, 0.7676) |
| 1000 | -0.08318 | -3.92518 | (0.6992, 0.7559) |
| 2500 | -0.08430 | -3.82622 | (0.6973, 0.7441) |
| 5000 | -0.08515 | -3.83708 | (0.6973, 0.7402) |
| 7500 | -0.08619 | -3.89263 | (0.6973, 0.7402) |

Table 6) The properties of the primary eddy, for right hand side aligned right triangle as a function of Reynolds number

| Re | $\psi$ | $\omega$ | $(x, y)$ |
|---|---|---|---|
| 100 | -0.06451 | -5.01902 | (0.4473, 0.8516) |
| 500 | -0.06065 | -5.73737 | (0.5469, 0.8496) |
| 1000 | -0.05306 | -7.02235 | (0.6094, 0.8691) |
| 1500 | -0.04765 | -8.20570 | (0.6582, 0.8848) |
| 2000 | -0.04353 | -9.32624 | (0.6953, 0.8965) |
| 2500 | -0.04019 | -10.4208 | (0.7227, 0.9043) |

Table 7) The properties of the primary eddy, for left hand side aligned right triangle as a function of Reynolds number